\documentclass[%
 reprint,
 amsmath,amssymb,
 aps,
]{revtex4-2}

\usepackage{graphicx}
\usepackage{subcaption}
\usepackage[justification=justified,singlelinecheck=false]{caption}
\usepackage{ragged2e}
\usepackage{lipsum}
\usepackage{dcolumn}
\usepackage{bm}
\usepackage[usenames]{color}
\usepackage[hidelinks,colorlinks=true]{hyperref}
\usepackage[mathlines]{lineno}
\usepackage{xcolor}
\usepackage{comment}
\usepackage{float}

\begin{document}

\preprint{APS/123-QED}

\title{\Large{\textcolor{teal}{ Rare Higgs Boson Decay into a Photon and a Z Boson in \\ Radiatively-Driven Natural Supersymmetry}}}

\author{E. A. Reyes R.}
\email{edilson.reyes@unipamplona.edu.co}
\author{C. A. Lopez A.}%
\author{O. R. Torrijo G.}%
\affiliation{ Departamento de Fisica - GOM, Universidad de Pamplona, \\ Pamplona - Norte de Santander, Colombia.}

\author{D. G. Melo P.}
\affiliation{ Department of Physics \& Astronomy, University of Pittsburgh,\\
Pittsburgh, PA 15260, USA.} 


\begin{abstract}
In this article, we study the rare decay process in which a Higgs boson decays into a $Z$ boson and a photon. In the first part of the paper, we analyze the Standard Model (SM) contributions to the corresponding decay width, including the full leading-order result, two-loop $O(\alpha_s)$ QCD corrections, and the recently reported two-loop $O(\alpha)$ electroweak corrections, evaluated under four different $\alpha$ renormalization schemes. The dependence on the Higgs boson mass is studied within the experimentally allowed range reported by the LHC. In the considered schemes, a non-negligible variation of about $4\%$ is found when the mass is varied within its current experimental uncertainty. In the second part of the paper, we analyze the leading-order contributions to the same process within the Minimal Supersymmetric Standard Model (MSSM). The spectrum of soft SUSY-breaking parameters and SUSY particle masses at the electroweak scale, which enter the computation of the one-loop amplitudes contributing to the decay width, is obtained by evolving the GUT-scale parameters of a Radiatively-Driven Natural Supersymmetry (RNS) model with non-universal Higgs boson masses. Variations of the RNS parameters can enhance the average SM prediction by up to $\sim20\%$, reaching a value of $\sim7.5~$keV, while still satisfying the Higgs boson mass constraint. However, this comes at the cost of allowing a moderately large fine-tuning parameter, with values exceeding $100$, thereby placing the model outside its most natural parameter region. The predicted decay width in the RNS scenario is closer to the recent ATLAS RUN 2 + 3 combined measurement than the average SM expectation.

\end{abstract}


\maketitle


\section{\label{sec:Introduction} Introduction \protect}

The high-precision measurements and global fits of electroweak (EW) and Higgs boson observables at the HL-LHC and future collider experiments offer a promising approach to search for signals of new physics that deviate from the SM predictions. In recent years, increasing attention has been devoted to precision observables that may exhibit potential deviations from SM predictions. The muon anomalous magnetic moment~\cite{B.Abi} and the $W$-boson mass~\cite{CDF,CMSmW} remain under close scrutiny due to their high experimental accuracy. Moreover, the recent evidence for the $h \rightarrow Z\gamma$ decay reported by both ATLAS and CMS at the LHC~\cite{HZgATLASCMS,ATLASlatest} has renewed interest in Higgs-sector observables. Although the current measurement is still affected by large statistical uncertainties, of about $40\%$ in the latest ATLAS analysis, the precision is expected to improve substantially at the HL-LHC~\cite{HL-LHC}, with a projected uncertainty of around $14\%$. Assessing the consistency of these results requires both enhanced experimental precision and refined theoretical predictions, while also motivating the exploration of possible extensions of the SM. Extensions including supersymmetry (SUSY) are well-known for enabling phenomenological analyses that can be experimentally tested. In particular, the muon g-2 anomaly~\cite{muontheory1,muontheory2,muontheory3,muontheory4,muontheory5,muontheory6} can be explained in non-universal SUSY models~\cite{Ellis2024}, where loop corrections involve electroweak sparticles with masses between $200$ and $700$~GeV, still allowed by current ATLAS and CMS searches~\cite{ATLAS1,CMS1,PDG}. However, this tension has disappeared~\cite{muonWP25,fnal25} in the latest SM prediction, where the HVP contribution is based on lattice methods rather than dispersive ones, which were affected by inconsistencies in the $e^+e^-\rightarrow\pi^+\pi^-$ data. In a similar manner, the $7\sigma$ deviation in the $W$ boson mass reported by CDF can be explained in the MSSM through loop corrections involving light EW/Higgs-sector sparticles and stop/sbottom contributions~\cite{MW1,MW2,MW3}. In specific parameter regions, the MSSM can also satisfy constraints from Dark Matter (DM) relic abundance~\cite{DM1}, DM direct detection~\cite{DM2, DM3, DM4}, and the LHC Higgs boson mass~\cite{MhExp}, while accommodating both the $g$–2 and $W$ mass anomalies~\cite{MW4}. In addition, Higgs boson precision measurements are expected to be conducted with high experimental accuracy at the FCC-ee~\cite{FCC1, FCC2, FCC3}. Therefore, uncertainties of Higgs properties will play an important role in the interpretation of the expected constraints on new physics. The Higgs boson mass ($M_h$) is currently measured with an experimental uncertainty of approximately $200$~MeV~\cite{DMhATLAS,DMhCMS} and it is expected to decrease to around $10$~MeV at FCC-ee. Its theoretical uncertainty is estimated based on the dependence of the three-loop QCD~\cite{HiggsMass1} and top-Yukawa~\cite{HiggsMass2} corrections on the renormalization scale ($Q$). When $Q$ is varied around the EW scale, $M_h$ changes around $50$ MeV, which gives a lower bound for the theoretical uncertainty. As pointed out in~\cite{FCC2, EWfit}, if the uncertainty is below $50$~MeV, the dependence on the Higgs mass results in a negligible error for the predictions of EW precision observables but leads to non-negligible parametric uncertainties of about $3\%$ in the Higgs decay processes $h \rightarrow Z\gamma$, $h \rightarrow WW$ and $h \rightarrow ZZ$. Specifically, this work focuses on the calculation of the decay width for the first process, where a Higgs boson decays into a $Z$ boson and a photon, analyzed in both the SM and MSSM frameworks. This rare decay can be regarded as a pseudo-observable extracted from the Dalitz process $h \to f\bar{f}\gamma$ by applying an invariant-mass cut around the $Z$ pole~\cite{Passarino:2013}. The latest combined analysis by ATLAS and CMS reports a branching ratio of $(3.4 \pm 1.1) \times 10^{-3}$, which is $\mu_s=2.2 \pm 0.7$ times larger than the SM prediction~\cite{HZgATLASCMS}, which represents a tension of $1.9\sigma$. Nevertheless, the most recent searches by ATLAS~\cite{ATLASlatest} observed a signal strength $\mu_s = 1.3^{+0.6}_{-0.5}$,  which is closer to the SM expectation. This is equivalent to an experimental branching ratio of $(2.03^{+0.94}_{-0.79}) \times 10^{-3}$ or a decay width of about $\Gamma_{Z\gamma}^{\text{exp}} \sim 8.3~\text{keV}$. Since this process occurs via loop diagrams, any observed excess could offer a valuable opportunity to probe potential contributions from physics beyond the SM (BSM). Extensions to the SM that aims to alleviate this discrepancy must preserve the predictions already in good agreement with experiment, while also reproducing the precise Higgs boson mass, which imposes some of the most stringent constraints on the parameter space. \\ \\ The paper is organized as follows. In Section~\ref{sec:SM} we study the decay processes $h \rightarrow Z\gamma$ in the context of the SM. The dependence of the $h \rightarrow Z\gamma$ decay width on the Higgs boson mass is analyzed. In Section~\ref{sec:RNS}, we present the technical details of the calculation of the $h \rightarrow Z\gamma$ decay width in the MSSM, using the Yennie gauge within the dimensional regularization scheme. We employ a non-universal supergravity model to obtain the numerical values of the masses and parameters involved in the computation. The required parameters at the EW scale are obtained by evolving the GUT scale inputs via renormalization group techniques. We then analyze the variation of the $h \rightarrow Z\gamma$ decay width and the MSSM prediction for the Higgs boson mass as functions of the RNS parameters. Finally we give our conclusions and perspectives. The main routines and data that produce the results presented in this paper are available at~\cite{data}.
 

\section{\label{sec:SM} The $h\rightarrow Z\gamma$ Decay in the SM \protect}
    \begin{figure}[]
    \centering
    \includegraphics[width=0.45\textwidth]{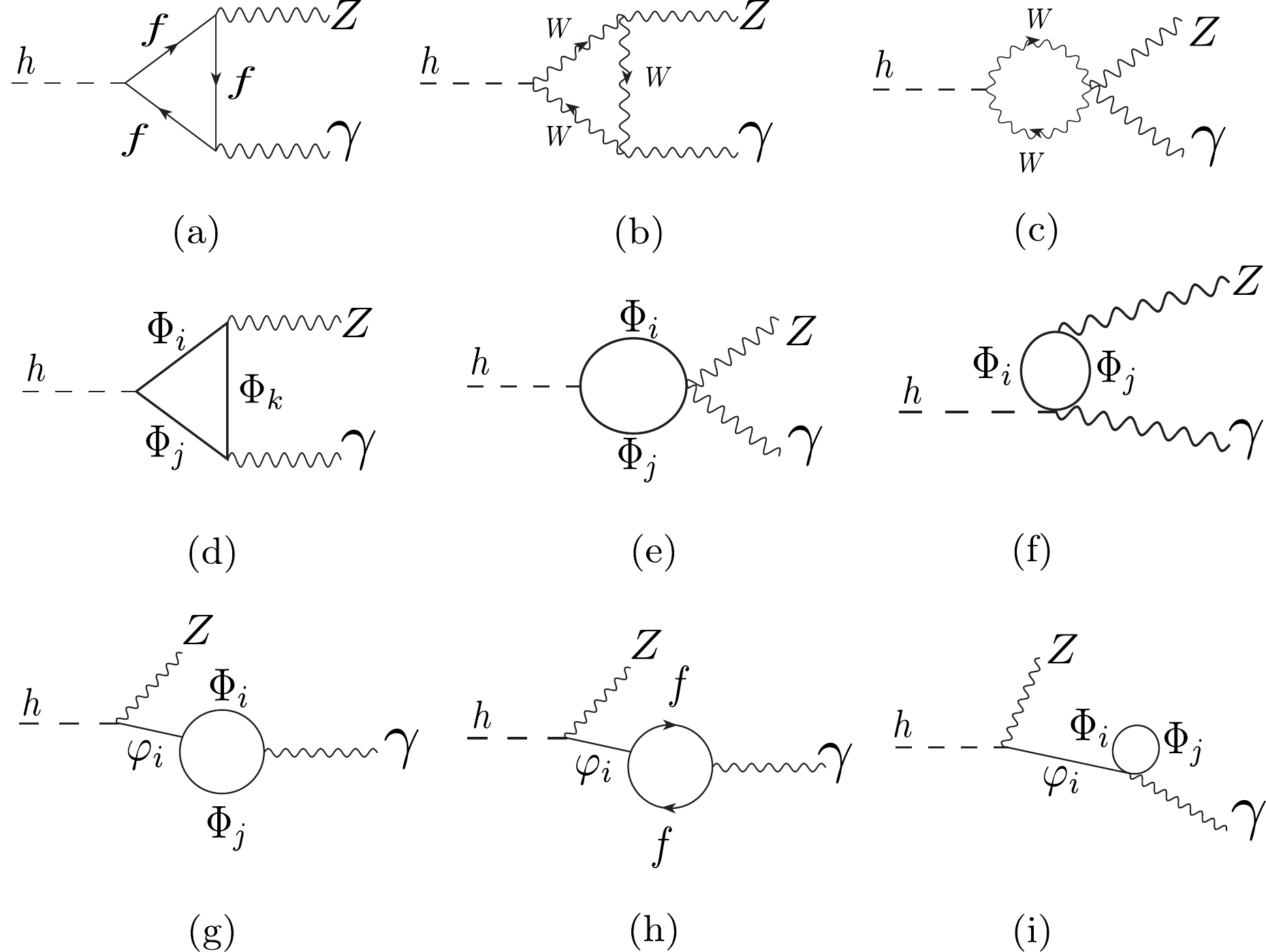}
    \caption{  \justifying  \small Feynman diagrams for the $h \rightarrow Z\gamma$ process in the SM at the one-loop level. Virtual fermions are denoted by $f$; $W$ represents the charged electroweak gauge boson; $\Phi_i$ stands for fields that include Goldstone bosons, W bosons and ghosts; and $\varphi_i$ in the non-1PI diagrams denotes either a virtual $Z$ boson or a neutral Goldstone boson.}
    \label{SMtopologies} 
\end{figure} 
We begin by reviewing the calculation of the $h \to Z\gamma$ decay width ($\Gamma_{Z\gamma}$) in the SM. Since there is no tree-level $hZ\gamma$ vertex in the SM Lagrangian, this process occurs only via radiative corrections, starting at the one-loop level. The relevant diagrams include loops involving fermions, EW gauge bosons, Goldstone bosons, ghosts, and other particles, as illustrated in FIG.~\ref{SMtopologies}. However, the dominant contribution arises from loops containing $W^{\pm}$ bosons~\cite{djouadiqcd, numerical}, shown in FIG.~\ref{SMtopologies}~(b) and (c). The absence of a $hZ\gamma$ vertex in the SM implies that the scattering amplitude must be finite~\cite{zdecay}, as there is no corresponding coupling that could absorb ultraviolet (UV) divergences through renormalization. In addition, non-1PI diagrams contribute to the amplitude, but only through mixing of the external photon leg with virtual $Z$ bosons via self-energy loop topologies (see FIG.~\ref{SMtopologies}~(g)-(i)). Similar diagrams involving the external $Z$ leg are forbidden due to the lack of any trilinear vertex involving the Higgs and the photon. \\ The fermion loop shown in FIG.~\ref{SMtopologies}~(a) is UV finite. In contrast, the contributions from gauge bosons, Goldstone bosons, and ghosts individually exhibit UV divergences arising from the gauge-dependent unphysical terms in their propagators. The finite amplitude at one-loop level is obtained only after summing all relevant diagrams, and its gauge dependence cancels out in the full result. \\ \\ In the unitary gauge, where Goldstone bosons and ghosts are absent, only the topologies shown in FIG.~\ref{SMtopologies}~(a)–(c) yield non-vanishing contributions. In contrast, the contributions from the diagrams in FIG.~\ref{SMtopologies}~(d)-(i) vanish. In particular, Feynman diagrams with the topology in FIG.~\ref{SMtopologies}~(g)-(h) are kinematically forbidden according to Cutkosky rules, as any unitary cut implies amplitudes with a single photon in the final state~\cite{dispersion} and for on-shell photons, there are no $Z$-$\gamma$ mixing propagators~\cite{anatomy}. It is worth mentioning that the finite part of the amplitude for $h \to Z\gamma$ and $h \to \gamma\gamma$ differs between the unitary gauge and the $R_\zeta$ gauge~\cite{2photonsdis,NPBgauge}. The issue originates from the gauge boson loop and the subtlety of taking the $\zeta \to \infty$ limit inside the momentum integral, leading to an apparent non-commutativity between these two operations. Gauge invariance can be restored using dispersion relations and the Goldstone Boson Equivalence Theorem (GBET), which allows the introduction of an additive constant so that the unitary gauge result matches that of the $R_\zeta$ gauge computed with dimensional regularization. Nevertheless, the problem of the apparent non-commutativity between the limit and the integration when transitioning from the $R_\zeta$ gauge to the unitary gauge remains unresolved. \\ \\ In this work, we reproduce the calculation of the SM decay width $\Gamma_{Z\gamma}$ in the $R_{\zeta}$ gauge at one-loop level. We have written a few \texttt{Wolfram Mathematica} scripts that uses \texttt{FeynCalc} \cite{FeynCalc1,FeynCalc2}, \texttt{FeynArts} \cite{FeynArts}, \texttt{FeynHelpers} \cite{FeynHelpers} and \texttt{Package-X} \cite{PackageX} to obtain the necessary expressions. The resulting amplitude for each individual diagram contributing to the decay width can be found in our GitHub repository. In general, it is possible to write the scattering amplitude for the $h\rightarrow Z\gamma$ process as~\cite{hzyrevisited}:
\begin{align} \label{desc}
i\mathcal{M}&=\mathcal{R}^{\mu\nu}\varepsilon_{\mu}^{\lambda_{1}*}\left(p_{1}\right)\varepsilon_{\nu}^{\lambda_{2}*}\left(p_{2}\right) \nonumber\\
&=\varepsilon_{\mu}^{\lambda_{1}*}\left(p_{1}\right)\varepsilon_{\nu}^{\lambda_{2}*}\left(p_{2}\right)\left[F_{0}g^{\mu\nu}+F_{1}p_{1}^{\mu}p_{2}^{\nu}+F_{2}p_{1}^{\mu}p_{1}^{\nu}\right. \nonumber\\
&\left. +F_{3}p_{2}^{\mu}p_{2}^{\nu}+F_{4}p_{2}^{\mu}p_{1}^{\nu}
+ F_5 \epsilon^{\mu\nu\alpha\beta}p_{1 \alpha}\ p_{2\beta}
\right]. 
\end{align}
The form factor $F_5$ gives a negligible contribution to the amplitude coming from chiral fermion loops. By using the Ward-Takahashi identity $\mathcal{R}_{\mu\nu}p_{2}^{\nu}=0$ together with the transversality condition of the Z boson polarization vector, $\varepsilon^{\lambda_{1}}\left(p_{1}\right)\cdot p_{1}=0$, the scattering amplitude can be succinctly written as:
\begin{equation}
i\mathcal{M}=\varepsilon_{\mu}^{\lambda_{1}*}\left(p_{1}\right)\varepsilon_{\nu}^{\lambda_{2}*}\left(p_{2}\right)F_{4}\left[-\left(p_{1}\cdot p_{2}\right)g^{\mu\nu}+p_{2}^{\mu}p_{1}^{\nu}\right].
\end{equation}
This allows for a simple expression of the corresponding decay width in terms of a single form factor:
\begin{equation}
\Gamma_{Z\gamma}^{\text{(SM)}}=\frac{M_{h}^{3}}{32\pi}\left(1-\frac{M_{Z}^{2}}{M_{h}^{2}}\right)^{3}\left|F_{4}\right|^{2}.
\label{eq:GammaSM}
\end{equation}
Each diagram contributing to $F_4$ is divided into a UV divergent term with a simple pole, as well as a finite part. After summing over all contributing amplitudes, the total simple pole term is given by
\begin{equation}
\Delta_{\varepsilon}^{SM}=\frac{i\pi^{2}e^{3}M_{W}\left(\zeta+3\right)}{2c_{W}^{3} M_Z^2 s_{W}^{2}} \left(M_W^2-c_W^2M_Z^2\right) \frac{1}{\varepsilon},\label{eq:divSM}
\end{equation}
where $\theta_W$ is the weak mixing angle, and we define $c_W=\cos\theta_W$ and $s_W=\sin\theta_W$. The term proportional to $M_W^2$ comes from the sum of all non-1PI diagrams (corresponding to the loop corrections represented in FIG.~\ref{SMtopologies}~(g)-(i)), while the remaining term proportional to $-c_W^2M_Z^2$ comes from the sum of 1PI diagrams depicted in FIG.~\ref{SMtopologies}~(b)-(f). Regardless of the choice of the gauge parameter $\zeta$, the coefficient of the simple pole $\varepsilon^{-1}$ vanishes in the full amplitude due to the mass relation between the $W$ and $Z$ bosons implied by custodial symmetry. Notably, in the Yennie gauge ($\zeta = -3$)~\cite{Yennie}, the divergent parts of both 1PI and non-1PI diagrams cancel separately, allowing their finite contributions to be extracted directly. \\ Using the Passarino-Veltman functions, we evaluate the finite part contributing to $F_{4}$, obtaining the same result as shown in~\cite{generaloneloop}. The dominant contributions come from the fermion and $W$ boson loops ~\cite{induced,hunter}: 
\begin{equation} \label{F4}
    F_4= \frac{e^3 s_W}{16\pi^2 M_W} \left( F_f + F_W + \dots \right).
\end{equation}
The $W$ boson loop contribution reads:
\begin{align}\label{fw}
 &F_{W} =-\frac{c_{W}}{s_{W}}\left\{ 4\left(3-t_{W}^{2}\right)I_{2}\left(\tau_{W},\lambda_{W}\right)\right.\nonumber \\
 & \left.+\left[\left(1+\frac{2}{\tau_{W}}\right)t_{W}^{2}-\left(5+\frac{2}{\tau_{W}}\right)\right]I_{1}\left(\tau_{W},\lambda_{W}\right)\right\},
\end{align}
while the fermion contribution is
\begin{align}\label{ff}
&\hspace{-0.1cm}F_{f}=\frac{-2qN_{c}}{s_{W}c_{W}}\left(T_{f}^{3}-2qs_{W}^{2}\right)\left[I_{1}\left(\tau_{f},\lambda_{f}\right)- I_{2}\left(\tau_{f},\lambda_{f}\right)\right],
\end{align}
where $t_W = \frac{s_W}{c_W}$, $\tau_i=\frac{4M_i^2}{M_h^2}$, $\lambda_i = \frac{4M_i^2}{M_Z^2}$,  $N_c$ is the color number, $T^3_f$ the weak isospin and $q$ the electric charge of the fermion. The functions $I_{1,2}$ which appear in expressions (\ref{fw}) and (\ref{ff}), are defined as:
\begin{align}
&I_{1}\left(a,b\right)  =\frac{ab}{2\left(a-b\right)}+\frac{a^{2}b^{2}}{2\left(a-b\right)^{2}}\left[f\left(a\right)-f\left(b\right)\right]\nonumber \\
 & +\frac{a^{2}b}{\left(a-b\right)^{2}}\left[g\left(a\right)-g\left(b\right)\right],\\
&I_{2}\left(a,b\right)  =-\frac{ab}{2\left(a-b\right)}\left[f\left(a\right)-f\left(b\right)\right].
\end{align}
In the kinematic region of interest, $4M_W^2>M_h^2$, the functions $f$ and $g$ are reduced to:
\begin{align}
f\left(x\right) & =\left[\sin^{-1}\left(\frac{1}{\sqrt{x}}\right)\right]^{2},\ \forall\ x\geq1, \\
g\left(x\right) & =\sqrt{x-1}\ \sin^{-1}\left(\frac{1}{\sqrt{x}}\right),\ \forall\ x\geq1 .
\end{align}
\begin{figure}[H]
    \centering
    \includegraphics[scale=0.85]{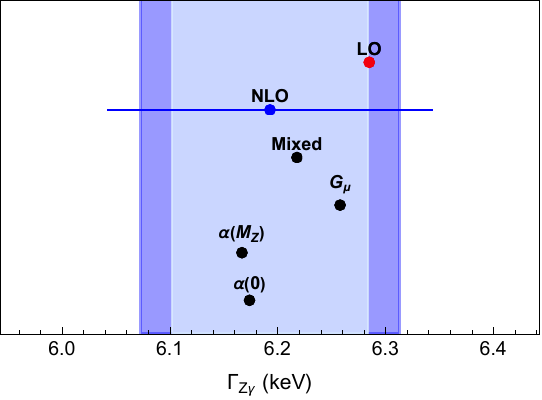}
    \caption{\small{The figure shows SM values for $\Gamma_{Z\gamma}$ at LO and NLO. Black dots represent NLO results in different renormalization schemes for $\alpha$. The red and blue dots represent the average of LO and NLO results among renormalization schemes. The error bar of the LO result is not shown. The blue (light blue) strip corresponds to the uncertainty of the NLO central value when only the dependence on the $M_h^2$ (the truncation of the perturbative series) is considered. }}
    \label{fig:gamNLO}
\end{figure}
\noindent Additionally, recent studies have quantified the signal–background interference between the Higgs-mediated and continuum $gg\rightarrow Z\gamma$ production processes, including NLO QCD corrections. The estimated interference has a destructive effect of about $-3\%$ on the total rate~\cite{EllisDjouadi}, which is small given the current experimental accuracy and does not significantly modify the SM prediction. Moreover, the next-to-leading order (NLO) electroweak contribution to the decay width has been computed and reported in references~\cite{EWNLO1,EWNLO2} using four different $\alpha$-renormalization schemes. The average LO and the average NLO values in the $\alpha$-schemes are represented in FIG.~\ref{fig:gamNLO}. In the $\alpha(0)$ scheme, all QED couplings in the three vertices of the LO diagrams are set to the fine-structure constant $\alpha(0)$ in the Thomson limit. In the $\alpha(M_Z)$ scheme, the two vertices not associated with photon emission are assigned $\alpha(M_Z)$, while the photon vertex remains at $\alpha(0)$. In the $G_\mu$ scheme, the same two vertices are instead assigned $\alpha_{G_\mu}$, with the photon vertex also kept at $\alpha(0)$. Finally, the democratic (Mixed) scheme, distributes the couplings among the three vertices using $\alpha_{G_\mu}$, $\alpha(M_Z)$, and $\alpha(0)$. The mean of these four values: $(6.19\pm0.15)$ keV, is also depicted with an error bar. This stands in contrast to the LO value: $(6.28\pm0.77)$ keV. There are two sources of uncertainty: i) the dependence on $M_h^2$ and ii) the truncation of the perturbative series. The first one is estimated via a linear approximation of $\Gamma$ as a function of $M_h^2$ and constitutes a $63\%$ of the variance of the width at NLO. To the second one we assign a variance equal to the maximum variation of $\Gamma_{Z\gamma}$ among the four schemes, thus obtaining the remaining $37\%$ of the variance of the NLO value. A lot of progress could be made in this direction with the projected uncertainties for the FCC--ee, which would place the truncation of the perturbative series well ahead as the leading component of uncertainty, as it would account for around $90\%$ of the variance of the width. This also highlights the need for new physics models where the Higgs mass is not an input, but a prediction, to improve their precision of $M_h$ in order to be tested against $\Gamma_{Z\gamma}$ and other electroweak observables.
\vspace{0.2cm}
\begin{figure}[H]
    \centering
    \includegraphics[scale=0.9]{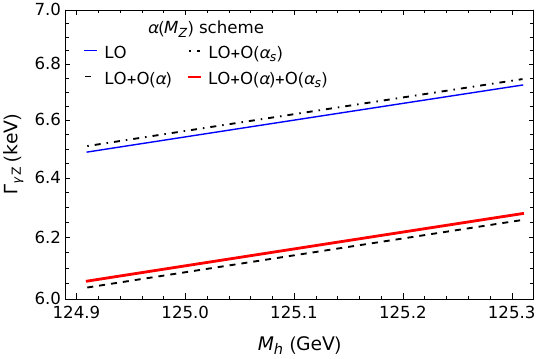}
    \caption{\small{The behavior of $\Gamma_{Z\gamma}$ within the $\alpha(M_{Z})$ scheme in the SM is observed across a range of Higgs masses, including the LO prediction and its EW and QCD corrections. }}
    \label{fig:AlpMZ}
\end{figure}
It is worth noting that all the four $\alpha$-schemes shows a similar variation with respect to the Higgs boson mass. The FIG.~\ref{fig:AlpMZ} displays the dependence of $\Gamma_{Z\gamma}$ on $M_h$ in the SM at different perturbative orders in the $\alpha(M_Z)$ scheme, which exhibits the biggest EW correction. Note that the QCD correction to the top-quark loop (dot-dashed black line), which is positive and amounts to only $0.3\ \%$ of the LO result (blue line), is smaller than the NLO electroweak correction (dashed black line), which is negative and contributes $13\ \%$ to the LO prediction. It is important to emphasize that the NLO electroweak corrections introduce noticeably different shifts to the LO predictions depending on the choice of the $\alpha$-scheme. In the $\alpha(M_Z)$ and $\alpha(0)$ schemes, the $\mathcal{O}(\alpha)$ corrections are roughly one order of magnitude larger than in the $G_\mu$ scheme. However, once the LO and NLO contributions are combined, the scheme dependence is significantly reduced, and the resulting predictions tend to converge to nearly the same value. This combined result is represented by the red solid line in FIG.~\ref{fig:AlpMZ} for the $\alpha(M_Z)$ scheme. A non-negligible variation is observed, from $\Gamma_{Z\gamma}=6.05$~keV to $\Gamma_{Z\gamma}=6.28$~keV, corresponding to an increase of about $3.8\%$, as the Higgs-boson mass spans the combined ATLAS/CMS measurement, $M_h=125.09\pm0.21$~GeV. The same relative increase is observed across the other schemes. \\ Given the latest combined results from ATLAS and CMS on the $h\rightarrow Z\gamma$ decay: $\mu_s=2.2\pm0.7$, the tension with the SM prediction stands at $1.9\sigma$ and the null hypothesis of compatibility with the SM has a p-value of approximately $6\%$~\cite{HZgATLASCMS,HZPCMS}. This analysis comes from the combination of  ATLAS' and CMS' respective datasets which had independently led to signal strengths of $2.0^{+1.0}_{-0.9}$ and $2.4\pm0.9$. However, the ATLAS collaboration recently released an analysis for data collected on 2022-2024, where the signal strength is measured to be $0.9^{+0.7}_{-0.6}$. This leaves the overall ATLAS result at $1.3^{+0.6}_{-0.5}$, which constitutes the most precise measurement of this channel to date. Considering these experimental results, there is no statistically significant tension between measurements and the SM. However, various models had been proposed to account for the mild tension before ATLAS' most recent results, such as: the presence of vector dark matter via the type II Seesaw mechanism~\cite{DMSeeSaw}, which adjusts to the observed data within $2\sigma$; a model incorporating charged Higgs doublets within the loop without affecting the $h\rightarrow \gamma\gamma$ process~\cite{DHiggsC}; and the existence of new quarks, where the CP-violating factor $h_{3}^{Z\gamma}$ would account for the excess in $\mu_s$~\cite{CPViolation}. Supersymmetric theories that adjust well to the current experimental context are discussed in the next section.


\begin{table*}[t]
\centering
\begin{tabular}{|l|c|c|c|c|c|}
\hline
\textbf{Model} & \textbf{Universality at GUT} & \textbf{Small $\mu$} & \textbf{Light Stops} & \textbf{$M_h = 125$ GeV} & \textbf{Naturalness} \\
\hline
mSUGRA         & yes     & no     & no         & large fine-tuning        & no \\
NS             & no      & yes    & yes        & moderate tuning           & partial \\
\textbf{RNS}   & partial & via RGE & moderate  & yes                       & yes \\
\hline
\end{tabular}
\caption{\small{Comparison between mSUGRA, Natural SUSY and Radiative Natural SUSY.}}
\label{tab:RNScomparison}
\end{table*}

\section{\label{sec:RNS} The $h\rightarrow Z\gamma$ decay in RNS} 

Detailed computations of the $h\rightarrow Z\gamma$ decay width have been carried out in different supersymmetric scenarios~\cite{hzgSUSY}. A wide range of phenomenological analyses provide SUSY spectra that modify the SM prediction of $\Gamma_{Z\gamma}$ while correctly reproducing the Higgs boson mass measured at the LHC. In the MSSM~\cite{MSSM1,MSSM2}, the net SUSY contribution to $\Gamma_{Z\gamma}$ can enhance the SM prediction by up to $30\%$ in certain scenarios~\cite{NHMSSM}. For large stop masses ($m_{\tilde{t}_j}$) above $600$~GeV, the radiative corrections to the predicted Higgs boson mass ($M_h$) can be large enough to yield ~$125$~GeV, as these corrections increase logarithmically with $m_{\tilde{t}_j}$, but this implies a large electroweak fine-tuning parameter ($\Delta_{EW}>30$). This model requires low values of third generation squark masses ($m_{\tilde{t}_j},\, m_{\tilde{b}_j} \leq 600$ ~GeV) in order to ensure a small value of $\Delta_{EW}$, as was discussed in~\cite{Baer2012,Baer2013}. As pointed out in these works, $\Delta_{EW}$ is the most conservative definition of a fine-tuning parameter. It is model independent, simple to calculate, and easily falsifiable, since requiring a low $\Delta_{EW}$ implies the existence of light Higgsinos with masses around $200$~GeV, which can be tested at a future $e^+ e^-$ collider. To reconcile the correct $M_h$ prediction with a small $\Delta_{EW}$, one possibility is to go beyond the MSSM by including additional multiplets, as in the case of the the Next-to-Minimal Supersymmetric Standard Model (NMSSM)~\cite{NextMSSM} where $\Gamma_{Z\gamma}$ can be at most 10\% higher than the SM result. Other possible extensions include the constrained MSSM (CMSSM)~\cite{CMSSM}, where the predictions of $\Gamma_{Z\gamma}$ are slightly below the SM result, or the nearly MSSM (nMSSM)~\cite{nMSSM} where the signal rate is significantly suppressed. Recent research indicates that the BLSSM~\cite{BLSSM}, the $\mu$ from $\nu$ MSSM ($\mu\nu$SSM)~\cite{uvSSM} and the non-holomorphic extension (NHSSM)~\cite{NHMSSM} could lead to sizable deviations from the SM predictions (even though the effects of the NHSSM couplings are negligible), bringing them closer to the experimental value. However, all these extensions involve two important problems that need to be taken into account. The requirement of a low $\Delta_{EW}$ strongly disfavors most of them, given that light third-generation squarks lead to a predicted branching fraction for the $b\rightarrow s \gamma$ decay that is significantly lower than the measured value~\cite{BaerJHEP2012}. Besides, the theoretical uncertainty in the Higgs boson mass within the minimal SUSY extension of the SM ($1-5$~GeV) is about one order of magnitude larger than the experimental precision achieved at the LHC, and this uncertainty increases even further in extensions of the MSSM~\cite{HiggsSUSY}. \\ An alternative to deal with these problems is given by models of Radiative Natural Supersymmetry (RNS), which can be realized within the MSSM avoiding the inclusion of additional exotic matter~\cite{Baer2013}. The RNS models are well motivated. They can induce low fine-tuning at the $3$\% level ($\Delta_{EW}$ $\sim 30$) for a large stop mixing parameter ($X_t$) with moderately heavy top squarks ($m_{\tilde{t}_j}\approx 1-5~$TeV), allowing for a $125$~GeV light Higgs boson while respecting LHC constraints on sparticle and gluino masses, and contraints from $B$ physics, especially the branching ratio for $b\rightarrow s \gamma$  decay. RNS is a refined framework that builds upon both mSUGRA (or CMSSM) and Natural SUSY (NS). In contrast to mSUGRA, which assumes strict universality of scalar masses ($m_0$), gaugino masses ($m_{\frac{1}{2}}$) and soft trilinear parameters ($A_0$) at the GUT scale ($\Lambda_{GUT}\approx 10^{16}~$GeV), typically leading to large fine-tuning, RNS allows for partial non-universality, especially in the Higgs scalar sector, where the masses $m_{H_1}^2$ and $m_{H_2}^2$ are chosen independently of $m_0$ at the GUT scale. The parameters $m_{H_1}^2$ and $ m_{H_2}^2$ are soft SUSY-violating mass terms associated with the Higgs doublets $H_1$ and $H_2$, which couples to up and down type quarks respectively in the MSSM (we follow the conventions of~\cite{HiggsParticles}). At the EW scale ($M_Z\approx 91~$GeV) $m_{H_1}^2$ plays a central role in the EW spontaneous symmetry breaking. For moderate $\tan\beta=t_{\beta}=v_2/v_1$ values (ratio of the VEVs of the two MSSM Higgs doublets) the minimization of the Higgs effective potential leads to the well-known relation:
\begin{equation}
\frac{M_Z^2}{2} \approx \frac{m_{H_2}^2 + \Sigma_2^2 - \left(m_{H_1}^2 + \Sigma_1^1 \right) t_{\beta}^2}{t_{\beta}^2 - 1} - \mu^2, \label{eq:ewsb}
\end{equation}
where $\Sigma_{j}^{j}$ are radiative corrections coming from the derivatives of the one-loop Higgs effective potential while $\mu$ is the mass scale of the Higgsinos. For large negative values of $X_t$, the correction $ \Sigma_{1}^{1}$ is suppressed and at the same time $M_h=125~$GeV is reached. This implies that natural electroweak symmetry breaking requires $m_{H_1}^2$ to be negative and not much larger in magnitude than $M_Z^2$. Otherwise, a large cancellation between $m_{H_1}^2$ and $\mu^2$ would be necessary, indicating fine-tuning. Natural SUSY improves naturalness by keeping only third-generation squarks light and lowering the $\mu$ parameter, but it often requires some degree of fine-tuning to reproduce the measured Higgs boson mass. RNS retains the key idea of a small $\mu$ but generates it naturally through the renormalization group evolution of $m_{H_1}^2$. Here, $m_{H_1}^2$ is allowed to be large and positive at the GUT scale, but it evolves to small negative values at the EW scale. This behavior enables a naturally small $\mu$ parameter and avoids significant fine-tuning while preserving proper electroweak symmetry breaking  and reproducing the correct Higgs mass value without the need for excessively light sparticles. A comparative summary of mSUGRA, NS, and RNS is shown in TABLE~\ref{tab:RNScomparison}. \\ \\ In our analysis, we estimate $\Gamma_{Z\gamma}$ within the MSSM using the low-energy values of the relevant parameters, such as $\mu$, $M_2$, $A_t$, $t_{\beta}$ and the stop masses, appearing in the one-loop amplitude. These parameters are obtained by solving the renormalization group equations (RGEs) starting from boundary conditions defined at $\Lambda_{GUT}$, as specified in the RNS framework. We implement RNS through the non-universal Higgs mass (NUHM) extension of the mSUGRA model, where $m_{H_1}^2$ and $m_{H_2}^2$ are replaced by the two weak scale parameters: $\mu$ and the mass of the CP-odd Higgs boson $m_A$, according with
\begin{equation}
    m_A^2 = m_{H_1}^2 + m_{H_2}^2 + 2\mu^2 \label{eq:mA},
\end{equation}
and the equation (\ref{eq:ewsb}). The model is then completely specified by the parameter set:  
\begin{equation}
    \mu, \, m_A, \, t_{\beta}, \, m_0, \, m_{\frac{1}{2}}, \, A_0. 
\end{equation}
The appropriate region of RNS parameters at $\Lambda_{GUT}$ is shown in TABLE~\ref{tab:RNSparam}. 
\begin{table}[H]
\centering
\begin{tabular}{|c|c|}
\hline
\textbf{RNS Parameter} & \textbf{Range at $\Lambda_{GUT}$} \\
\hline
$m_0$           & $2 - 8\ \mathrm{TeV}$ \\
$|A_0|$     & $2.0 - 16\ \mathrm{TeV}$  \\
$m_{\frac{1}{2}}$       & $0.3 - 1.5\ \mathrm{TeV}$ \\
$t_{\beta}$     & $10 - 20$ \\
$\mu$           & $0.1 - 0.35\ \mathrm{TeV}$ \\
$m_A$           & $0.15 - 1.5\ \mathrm{TeV}$ \\
\hline
\end{tabular}
\caption{\small{Typical parameter ranges at the GUT scale for RNS models that can reproduce $M_h \approx 125\ \mathrm{GeV}$ and maintain $\Delta_{\mathrm{EW}}$ small.}}
\label{tab:RNSparam}
\end{table}
In this case, $m_0$ represents the unified mass of the three generations of squark and slepton particles. The required mass spectrum and coupling values at the EW scale are obtained radiatively with the help of \texttt{Isasugra}~\cite{Isasugra} which is a SUSY spectrum generator included in the Monte Carlo program \texttt{IsaJet~7.91}~\cite{Isajet}. The Higgs boson mass is calculated using \texttt{FeynHiggs 2.19.0}~\cite{FeynHiggs}, with the flags set to include the full one-loop corrections, two-loop corrections at order $\alpha_s\alpha_t$, $\alpha_s\alpha_b$, $\alpha_t\alpha_b$, $\alpha_t^2$ and NNLL resummation of large logarithms in the MSSM with real parameters. Moreover, the one-loop expression of the $\Gamma_{Z\gamma}$ was independently computed by our group in the MSSM using the following specifications.   

\subsection{\label{sec:FDC} Diagrammatic Calculation of $\Gamma_{Z\gamma}$}

\begin{figure}[H]
    \centering
    \includegraphics[width=0.45\textwidth]{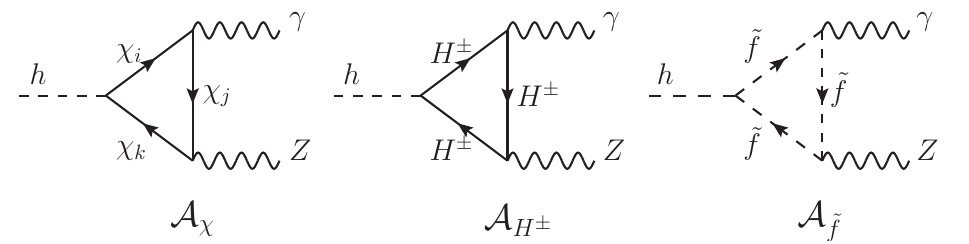}
    \caption{\small{Main diagrams contributing to $\mathcal{A}_{\text{SUSY}}$ in the $h\rightarrow Z\gamma$ decay width. The internal solid lines represent charginos or charged Higgs bosons, while the internal dashed lines represent sfermion fields.}}
    \label{fig:hZgSUSY}
\end{figure}

Besides the SUSY counterparts of the virtual loops depicted in FIG.~\ref{SMtopologies}, the MSSM includes new one-loop diagrams involving charged Higgs bosons~($H^{\pm}$), charginos~($\chi_j$) and sfermions~($\tilde{f}_j$). The general form of the one-loop decay width is given by
\begin{equation}
\Gamma_{Z\gamma}^{(\text{MSSM})} = \frac{\alpha G_F^2 M_W^2 M_h^3}{64 \pi^4} \left(1 - \frac{M_Z^2}{M_h^2} \right)^3 \left| \mathcal{A}_{\text{SM}} + \mathcal{A}_{\text{SUSY}} \right|^2, \label{eq:GammaMSSM}
\end{equation}
where $\mathcal{A}_{\text{SM}}$ includes the contributions from SM-like particles, and $\mathcal{A}_{\text{SUSY}}$ encapsulates diagram's amplitudes with the additional degrees of freedom proper of the MSSM. To draw all the diagrams and generate their corresponding amplitudes in D-dimensions, using dimensional regularization, we used \texttt{FeynCalc} together with \texttt{FeynArts}. We also employed \texttt{FeynHelpers}, which connects \texttt{FeynCalc} with \texttt{Package-X}, to obtain the divergent and finite parts of each diagram in terms of a basis of the Passarino-Veltman functions, and to perform any required algebraic manipulation on the amplitudes. It is worth mention that $\mathcal{A}_{\text{SM}}$ in eq. (\ref{eq:GammaMSSM}) is not the same SM amplitude $F_4$ of eq. (\ref{eq:GammaSM}) as in the MSSM all the cubic couplings including an external Higgs boson line differs from the corresponding SM ones by a factor that is function of the $(h, \, H)$ mixing angle $\alpha$ and the angle $\beta$. As a consequence, the divergent part of the MSSM one-loop scattering amplitude in the $R_{\zeta}$ gauge, differs from $\Delta_{\varepsilon}^{\text{SM}}$ of eq. (\ref{eq:divSM}) just by a factor that depends on these angles,
\begin{equation}
\Delta_{\varepsilon}^{\text{MSSM}}= \Delta_{\varepsilon}^{\text{SM}}\, \sin(\beta-\alpha). \label{divMSSM}
\end{equation} 
The UV divergences arise solely from the 1PI and non-1PI diagrams in $\mathcal{A}_{\text{SM}}$. The SUSY contribution, $\mathcal{A}_{\text{SUSY}}$, is finite, even though each individual diagram involving $\chi_j$, $H^{\pm}$ or $\tilde{f}_j$ contains a simple pole. This finiteness comes from cancellations among all diagrams with the same internal particle content. For instance, in the case of charginos, there are eight 1PI triangle diagrams (see $\mathcal{A}_{\chi}$ in FIG.~\ref{fig:hZgSUSY}) whose summed Feynman amplitudes yield a finite result. 
As the decay width is gauge invariant, we chose again the Yennie gauge to directly get rid of the divergent term $\Delta_{\varepsilon}^{\text{MSSM}}$ and obtain the corresponding finite expression. In the MSSM, the relevant contributions to $\Gamma_{Z\gamma}$ arise from loops involving the $W$ boson, the top quark, charginos and, in certain regions of parameter space, the charged Higgs bosons and third-generation sfermions, particularly staus~($\tilde{\tau}$) and stops~($\tilde{t}$), can provide sizable effects (see FIG. \ref{fig:hZgSUSY}). \\ As in the SM, the dominant contribution comes from loop diagrams with internal $W$ bosons.
Its amplitude differs from the SM one by an overall factor of $\sin(\beta - \alpha)$. In the decoupling limit, where $m_A \gg M_Z$, a regime typically satisfied in RNS, the mixing angle approaches $\sin(\beta - \alpha) \to 1$, and the MSSM contribution becomes indistinguishable from that of the SM. Although the chargino-induced amplitude does not exceed the magnitude of the $W$ boson contribution, it can surpass that of the top quark and remain phenomenologically relevant. This is especially true when the chargino spectrum is characterized by a light Higgsino-like state. In such cases, chargino loops can interfere constructively or destructively with the SM-like contributions. In the context of RNS, the parameter $\mu$ is small and leads to light higgsinos that mix with winos of mass $M_2\sim 0.3-1$~TeV to produce relatively low chargino masses ($m_{\chi_1}\sim 100-350~\mathrm{GeV}, \, m_{\chi_2}\sim0.5-1.5~\mathrm{TeV}$) obtained after diagonalizing the mass matrix: 
\begin{equation}
\mathcal{M}_C =
\begin{pmatrix}
M_2 & \sqrt{2} M_W \sin\beta \\
\sqrt{2} M_W \cos\beta & \mu
\end{pmatrix}.
\end{equation}
In particular, the chargino loop amplitude depends on the inverse of these low chargino masses and non-negligible couplings to the Higgs and gauge bosons, which is why this contribution dominates in $\mathcal{A}_{\text{SUSY}}$. It can be schematically expressed as:
\begin{equation}
\mathcal{A}_{\chi} \sim \sum_{i=1}^{2} \frac{1}{m_{\chi_i}} \left[ C_{L}^{(i)} F_L(x_i, y_i) + C_{R}^{(i)} F_R(x_i, y_i) \right],
\label{eq:Achi}
\end{equation}
where \( C_{L,R}^{(i)} \) are effective couplings involving the $h\chi_i\chi_j$ and $Z\chi_i\chi_j$ vertices, and the loop functions $F_{L,R}$ depend on $x_i = m_{\chi_i}^2 / M_h^2$ and $y_i = m_{\chi_i}^2 / M_Z^2$. This amplitude is also sensitive to the parameter $t_{\beta}$, which appears in both the masses and mixing of the charginos. \\ The amplitude arising from charged Higgs bosons has the form:
\begin{equation}
    \mathcal{A}_{H^\pm} \sim \frac{g M_W}{16\pi^2} \frac{\cos(\beta - \alpha)}{m_{H^\pm}^2} \, F_{H^\pm}(z),
    \label{eq:AH+}
\end{equation}
where $z = M_h^2 / (4 m_{H^\pm}^2)$ and $F_{H^\pm}(z)$ is a loop function. The charged Higgs mass $m_{H^\pm}$ is approximately related to the CP-odd Higgs mass $m_A$ via:
\begin{equation}
m_{H^\pm}^2 \approx m_A^2 + M_W^2,
\end{equation}
and lies in the $0.15$ - $1.5~\mathrm{TeV}$ range in RNS, making this contribution generally subdominant. \\ Scalar superpartners of fermions contribute through amplitudes of the form:
\begin{equation}
    \mathcal{A}_{\tilde{f}} \sim \frac{g^2 m_f^2}{m_{\tilde{f}}^2} \, N_C Q_f (T_3^f - Q_f \sin^2\theta_W) \, F_{\tilde{f}}(x_f),
    \label{eq:Asf}
\end{equation}
where $m_f$ is the mass of the corresponding fermion, $m_{\tilde{f}}$ is the sfermion mass, $N_C$ is the color factor~($3$ for squarks, $1$ for sleptons), $Q_f$ is the electric charge, and $T_3^f$ the weak isospin. The function $F_{\tilde{f}}(x_f)$, with $x_f = M_h^2/4 m_{\tilde{f}}^2$, captures the loop dynamics. Stops typically provide a subdominant sfermion contribution due to the relatively large stop mass. The stau contribution can be enhanced in scenarios with large  $t_{\beta}$ and significant stau mixing. In the RNS framework, while the stau-induced amplitude may become relevant for $t_{\beta}$ values approaching 20, it remains subleading compared to the chargino contribution. The explicit expressions for the amplitudes $\mathcal{A}_{j}$ and the loop functions $F_{i}$ can be consulted in our GitHub repository. 

\begin{figure*}[!htbp]
  \centering  
  \begin{subfigure}[c]{0.48\textwidth}
    \centering    \includegraphics[width=\linewidth]{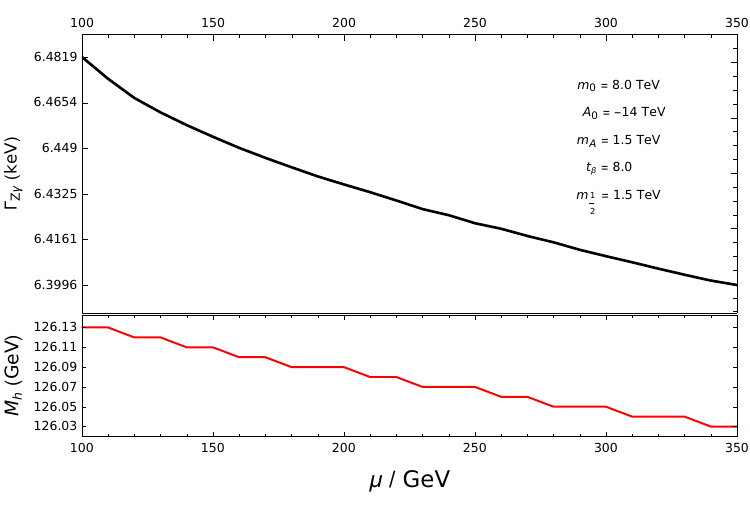}
    \caption{ }
    \label{fig:a}
  \end{subfigure}
  \begin{subfigure}[c]{0.48\textwidth}
    \centering    \includegraphics[width=\linewidth]{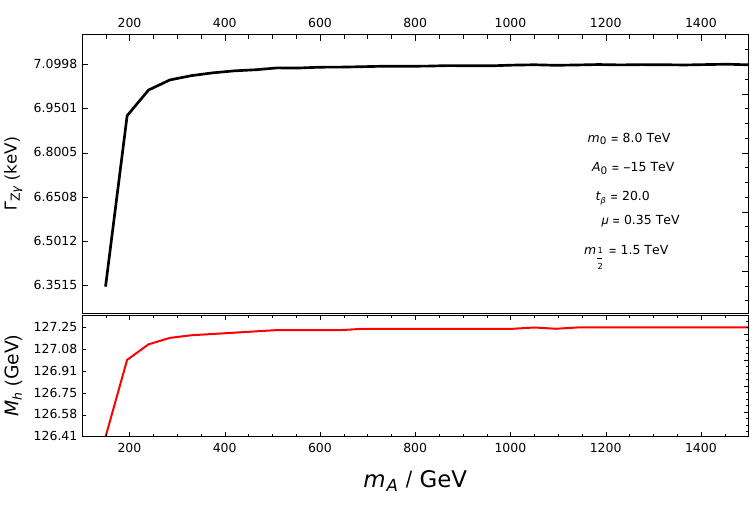}
        \caption{ }
        \label{fig:b}
  \end{subfigure}
   \begin{subfigure}[c]{0.47\textwidth}
    \centering
    \includegraphics[width=\linewidth]{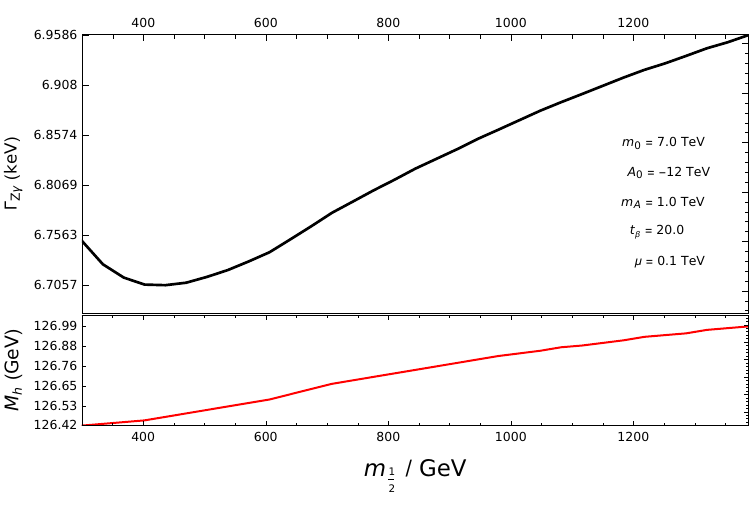}
        \caption{ }
        \label{fig:c}
  \end{subfigure}
  \begin{subfigure}[c]{0.48\textwidth}
    \centering
    \includegraphics[width=\linewidth]{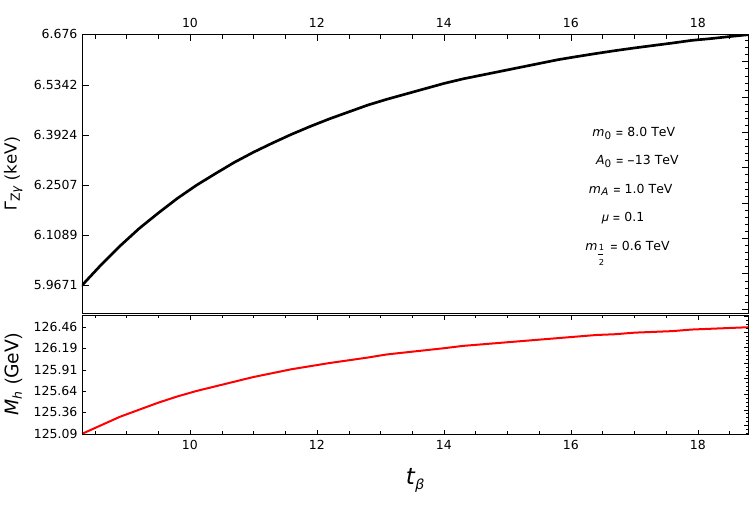}
        \caption{ }
        \label{fig:d}
  \end{subfigure}
   \caption{  \justifying \small{Variation of the Higgs decay width $\Gamma_{Z\gamma}$ and the MSSM Higgs boson mass $M_h$ as a function of the following input RNS parameters. (a) Higgsino mass, $\mu$, (b) CP-odd Higgs boson mass, $m_{A}$, (c) grand unified gaugino mass, $m_{\frac{1}{2}}$, and (d) ratio of the vevs of the two Higgs doublets, $t_{\beta}$. All of them evaluated at the GUT scale. }}
    \label{fig:fourplots}
\end{figure*}

\subsection{\label{sec:Numerical} Numerical Analysis}

\begin{figure}[h]
    \centering
    \includegraphics[width=0.45\textwidth]{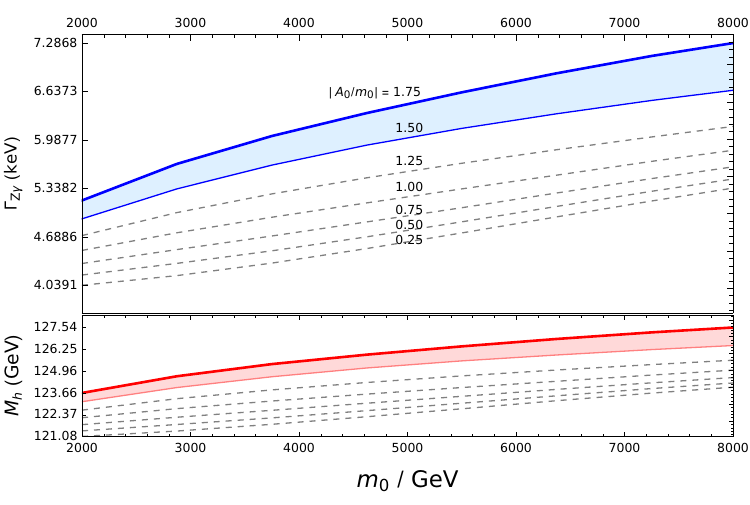}
    \caption{  \justifying \small{Variation of the $h\rightarrow Z\gamma$ decay width as a function of $m_0$ for different values of the ratio $|A_0/m_0|$.}}
    \label{fig:m0plot}
\end{figure}

\begin{figure}[h]
    \centering
    \includegraphics[width=0.45\textwidth]{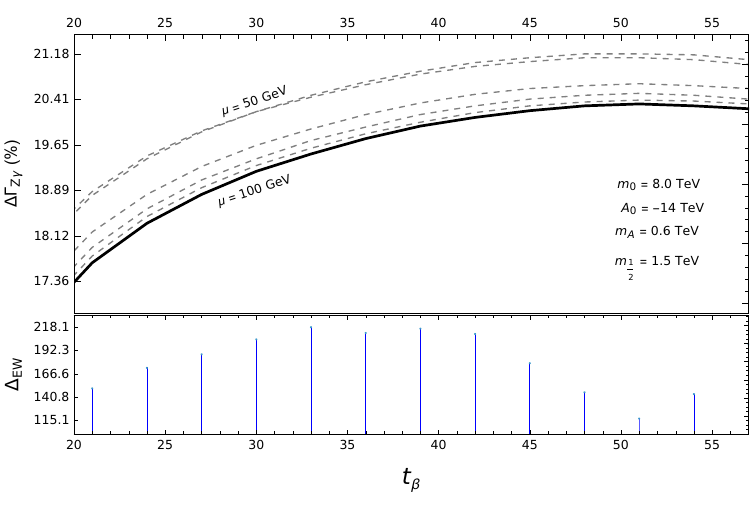}
    \caption{ 
        \justifying \small{Upper: Percentage difference between the RNS prediction and the NLO average of $\Gamma_{Z\gamma}$ in the SM as a function of $\tan \beta$. Lower: Electroweak fine-tuning parameter for different $\tan \beta$ values. }
       }
    \label{fig:ftplot}
\end{figure}

Accurate evaluation of the contributions in eqs. (\ref{eq:Achi}), (\ref{eq:AH+}) and (\ref{eq:Asf}) requires computing the loop functions $F_L$, $F_R$, $F_{H^\pm}$ and $F_{\tilde{f}}$ in terms of the masses $m_{\chi_j}$, $m_{H^{\pm}}$ and $m_{\tilde{f}}$ evaluated at the EW scale. Those values are obtained through the RGEs implemented in \texttt{IsaJet}, with the boundary conditions defined at $\Lambda_{GUT}=1.5\times10^{16}$~GeV in terms of the RNS inputs $\mu$, $m_A$, $m_{\frac{1}{2}}$, $m_0$, $A_0$ and $t_{\beta}$. The prediction of the $h\rightarrow Z\gamma$ decay width depends on the precise election of such a boundary condition. For this reason we study the behavior of $\Gamma_{Z\gamma}$ as a function of the RNS parameters in the ranges defined in TABLE~\ref{tab:RNSparam}. \\ In FIG.~\ref{fig:fourplots} we show the variation of the Higgs decay width $\Gamma_{Z\gamma}$ (black solid line) and the predicted three-loop MSSM Higgs boson mass $M_h$ (red solid line) as a function of the RNS inputs: (a)~$\mu$, (b)~$m_A$, (c)~$m_{\frac{1}{2}}$ and (d)~$t_{\beta}$. In these plots, each of the RNS parameters is varied one at a time, while the others are kept fixed at the values reported in each graph. Besides, we only admit RNS points that constrain the $M_h$-prediction to lie within the range $M_h=125.1\pm2.0$~GeV. It is noteworthy that, as shown in FIG.~\ref{fig:a}, the numerical prediction for $\Gamma_{Z\gamma}$ increases as the value of $\mu$ decreases, reaching its maximum for $\mu=100$~GeV. This behavior favors scenarios with low EW fine-tuning and a light Higgsino spectrum, which can be accessible at near-future collider experiments. On the other hand, as shown in the remaining plots, when $m_A$, $m_{\frac{1}{2}}$ and $t_{\beta}$ are increased, the Higgs decay width also increases, showing a strong dependence on the parameter $t_{\beta}$ (FIG.~\ref{fig:d}) which changes the value of $\Gamma_{Z\gamma}$ from $~6$~keV to $~7$~keV (around $17\%$) when $t_{\beta}$ evolves from $8$ to $20$. Values of $\tan\beta \lesssim 8$ are generally disfavored. In this regime, the top Yukawa coupling becomes very large, potentially hitting a Landau pole below the GUT scale and leading to a breakdown of perturbativity. Furthermore, low $\tan\beta$ tends to suppress the predicted Higgs mass, making it difficult to achieve the observed value of $125$~GeV without relying on extreme stop mixing. Additionally, note that the dependence on $m_A$ (FIG.~\ref{fig:b}) is negligible for values above $300$~GeV, where the curve remains essentially flat reaching a plateau, but it decreases rapidly below this threshold. This behavior once again favors a small $\Delta_{\text{EW}}$, since larger values of $m_A$ typically lead to increased electroweak fine-tuning. Finally, there is only a moderate dependence on $m_{\frac{1}{2}}$ (FIG.~\ref{fig:c}), the decay width increases by approximately $3.7\%$ as $m_{\frac{1}{2}}$ is varied from $500$ to $1500$~GeV. \\ The dependence of the Higgs decay width on $A_0$ and $m_0$ must be carried out carefully. In the RNS framework, the ratio $|A_0 / m_0|$ plays a critical role in determining the viability of the scalar spectrum. Large values of this ratio can enhance stop mixing, which is beneficial for raising the Higgs mass to the observed value near $125$~GeV. However, when $|A_0 / m_0|$ becomes too large, typically beyond the range $ 1.5 \lesssim |A_0 / m_0| \lesssim 2.0 $, the RG evolution of scalar soft masses can lead to tachyonic states. These usually appear in the third-generation sfermion sector, particularly in the lighter stop $\tilde{t}_1$ or stau $\tilde{\tau}_1$, where the large trilinear terms induce sizable off-diagonal mixing that can drive one of the squared mass eigenvalues negative. In some cases, tachyonic masses may also arise in the Higgs scalar sector, signaling an instability in the electroweak vacuum. \texttt{IsaJet} detects these issues during spectrum generation and exclude such parameter points as physically inconsistent. Therefore, while moderate values of $|A_0 / m_0|$ are necessary to achieve realistic scalar masses in RNS, excessively large values must be avoided to ensure a viable and stable scalar spectrum. Having this in mind, the variation of $\Gamma_{Z\gamma}$ with respect to the ratio $|A_0/m_0|$ is shown in FIG.~\ref{fig:m0plot}. For this analysis, we have used the RNS point that yields the largest value of $\Gamma_{Z\gamma}$ according to FIG.~\ref{fig:fourplots}, namely: $\mu=100$~GeV, $m_A=1500$~GeV, $m_{\frac{1}{2}}=1400$~GeV and $t_{\beta}=20$. The blue band highlights the points that are consistent with the accepted range of the $M_h$-prediction, which is represented by a red band in the lower panel of the graph. Consequently, points with $|A_0/m_0|<1.5$ (indicated by dashed gray lines) are excluded. Moreover, larger values of both $m_0$ and $|A_0/m_0|$ lead to higher predictions for $\Gamma_{Z\gamma}$. In particular, for $|A_0/m_0|=1.75$ and $m_0=8000$~GeV, the decay width can reach $\Gamma_{Z\gamma}\approx 7.3$~keV, which is $20.7\%$ higher that the lowest SM prediction in the $\alpha(M_Z)$ scheme ($\Gamma_{Z\gamma}=6.05$~keV) shown in FIG.~\ref{fig:AlpMZ}, and $17.9\%$ higher when compared with the NLO average ($\Gamma_{Z\gamma}=6.19$~keV) represented by the blue point in FIG.~\ref{fig:gamNLO}. The ranges of the RNS parameters shown in figures~\ref{fig:fourplots} and \ref{fig:m0plot} were constrained in order to balance the Higgs mass prediction with low fine-tuning. When the RNS parameters are closer to their maximal considered values, the fine-tuning parameter increases while maintaining the correct Higgs boson mass, reaching values of $\Delta_{EW}$ above $60$, yet it always remains below $100$. The numerical value of the predicted $\Gamma_{Z\gamma}$ can be slightly increased by taking larger values of $\tan\beta$ and smaller values of the $\mu$ parameter. 
However, the following important considerations must be taken into account. (i) Values of $\mu$ below $100$~GeV are strongly disfavored due to experimental constraints. In this regime, the lightest chargino and neutralino become predominantly Higgsino-like and acquire masses close to $\mu$. Current limits from LEP and LHC direct searches require $m_{\tilde{\chi}_1^\pm} \gtrsim 100$~GeV in typical MSSM scenarios. (ii) Although $\tan\beta$ can be greater than $20$, it should remain below $60$ in order to avoid issues with perturbativity and electroweak vacuum stability. Values of $\tan\beta \gtrsim 55$ typically lead to the appearance of tachyonic staus due to large left-right mixing terms driven by the enhanced tau Yukawa coupling. This can also result in charge-breaking minima in the scalar potential. Additionally, for $\tan\beta \gtrsim 60$, numerical instabilities commonly arise in \texttt{IsaJet} and \texttt{FeynHiggs}, which signal the onset of unphysical or unreliable solutions. (iii) Increasing $\tan\beta$ also leads in general to an increase in fine-tuning. \\ In FIG.~\ref{fig:ftplot}, we display the percentage increase of the MSSM-RNS prediction for the decay width of $h\rightarrow Z\gamma$, computed relative to the SM NLO average value of $\Gamma_{Z\gamma}$ (see FIG.~\ref{fig:gamNLO}). The lower panel also displays the value of the EW fine-tuning parameter $\Delta_{EW}$ for different values of $\tan \beta$. In both plots, we scan over the range $20\leq t_\beta \leq 60$. As can be seen from the black solid line, the RNS prediction is consistently higher than the SM average value across the scanned range, reaching a maximum difference of $20.34\%$ (corresponding to $\Gamma_{Z\gamma}\sim7.5~$keV) for $t_{\beta}$ approaching $57$. Tachyonic states appear for $t_\beta > 57$. The precise RNS point that maximizes the prediction of $\Gamma_{Z\gamma}$ is: $\mu=100~$GeV, $m_0=8.0~$TeV, $A_0=-14~$TeV, $m_A=0.6~$TeV, $m_{\frac{1}{2}}=1.5~$TeV and $t_{\beta}=51$. Values of $\mu$ below $100~$GeV (dashed gray lines) can further increase the percentage difference, but are excluded by current LHC searches. Finally, in order to obtain the maximum allowed value of the decay width prediction in the RNS scenario, it is necessary to allow the fine-tuning parameter to exceed $100$, as can be seen from the lower panel of FIG~\ref{fig:ftplot}. 
Although the RNS framework is designed to minimize electroweak fine-tuning, some tension arises when attempting to maximize the rare decay width $\Gamma_{Z\gamma}$, particularly in the large $\tan\beta$ regime. Achieving the highest predicted value requires pushing $\Delta_{\mathrm{EW}}$ beyond the nominal naturalness threshold. For instance, at $t_\beta = 51$, we obtain $\Delta_{EW} = 117$. This corresponds to a fine-tuning level of approximately $1/117$, or about $0.85\%$, indicating that a delicate cancellation among parameters is needed to stabilize the electroweak scale. While this is significantly less natural than the typical RNS benchmark range $\Delta_{EW} \lesssim 30$ (corresponding to $\gtrsim 3\%$ naturalness), it remains substantially better than in other SUSY models, such as PeV or minisplit supersymmetry~\cite{Baer2025}, where fine-tuning often exceeds $0.2\%$ (i.e., $\Delta_{EW} \gtrsim 500$). Therefore, while the point with $\Delta_{EW} = 117$ lies outside the natural region of RNS, it may still be considered moderately fine-tuned and phenomenologically acceptable, especially if the enhancement in the decay width leads to deviations from the SM prediction that are closer to the latest combined ATLAS measurement.

\section{\label{sec:Conclusions} Conclusions}

In this work we have revisited the amplitude for the $h\rightarrow Z\gamma$ decay width in the SM at LO. We have computed the full one-loop contribution in the $R_{\zeta}$ gauge in terms of the Passarino-Veltmann master integrals using dimensional regularization. We found that the spurious UV divergences, emerging in the regularization procedure, can be efficiently removed using the Yennie gauge, $\zeta = -3$. The finite contribution is gauge independent. Additionally, we analyzed the dependence of the decay width on the Higgs boson mass at both LO and NLO in four different renormalization schemes. The resulting decay widths are found to be on the order of keV, with a mean value of $6.19\pm0.15~$keV, showing a non-negligible variation of about $4\%$ in the four $\alpha$-schemes when the Higgs boson mass is varied over the range of its experimental measurement. We note that although the uncertainty in the decay width is reduced at NLO, its central value deviates more from the experimental result than the LO prediction because the two-loop EW corrections are negative. This motivates further work aimed at computing the missing NNLO contributions from EW and mixed QCD/EW corrections. In addition, we have calculated the MSSM corrections to $\Gamma_{Z\gamma}$ which come from loops with chargino, charged Higgs boson and sfermion particles. The numerical values of the corresponding sparticle masses at the EW scale were obtained by evolving the parameters of a supergravity model with non-universal Higgs boson masses (namely, RNS) from the GUT scale using the Rernormalization Group approach. The behavior of the $h\rightarrow Z\gamma$ decay width as a function of the RNS input parameters at $\Lambda_{GUT}$ was studied. We found that imposing in the RGEs the boundary condition where $\mu=100$~GeV, $m_{\frac{1}{2}}=1500$~GeV, $m_A=600$~GeV, $t_{\beta}=51$ and $m_0=8000$~GeV with $|A_0/m_0|=1.75$, the value of $\Gamma_{Z\gamma}$ increases up to about $7.5~$keV which is $\sim20\%$ higher than the average NLO prediction in the SM, while preserving the Higgs mass value inside the range $M_h=125\pm 2~$GeV and a moderately large fine-tuning parameter, $\Delta_{EW}=117$. Interestingly, the RNS prediction aligns well with the latest ATLAS result of $\sim 8.3~$keV and is even closer than the SM estimate of $\sim 6.2~$keV, suggesting a slightly improved description of this rare process within current uncertainties. Looking ahead, a possible enhancement of about $10$--$20\%$ in the partial width predicted within the RNS framework could be challenging but potentially testable at future colliders. At the HL-LHC, such a deviation would be close to the projected experimental precision on this channel, around $14\%$ for the combined ATLAS and CMS analyses. In contrast, lepton colliders such as the ILC~\cite{ILCwhitepaper} and FCC-ee~\cite{FCCeeHiggs} would offer much higher sensitivity thanks to their cleaner experimental environments and large Higgs production rates. In particular, the FCC-ee, expected to collect $\mathcal{O}(10^6)$ $Zh$ events, could reach percent-level accuracy in rare Higgs decay measurements, making a $10–20\%$ enhancement in $\Gamma_{Z\gamma}$ clearly observable. Finally, it is not possible to further increase the prediction, as important constraints forbid lowering the $\mu$ parameter below $100~$GeV and raising $\tan\beta$ beyond $60$. Our results show that enhancing the decay width $\Gamma_{Z\gamma}$ within the RNS framework, particularly in the large $\tan\beta$ regime, requires electroweak fine-tuning parameters exceeding $\Delta_{EW} \sim 100$. While such values are still more natural than in typical MSSM scenarios, they lie outside the preferred natural region of RNS defined by $\Delta_{EW} \lesssim 30$. This indicates that the requirement of maximizing the decay width prediction comes at the cost of increased fine-tuning, and thus reveals an intrinsic tension between naturalness and observable deviations from the Higgs properties in the SM. 

\section*{Acknowledgments}

We would like to thank to Professor Wen-Long Sang (SWU) for useful discussions about the technical aspects of his computation of the two-loop electroweak corrections to the $h\rightarrow Z\gamma$ decay width in the SM and for kindly sharing the data obtained from his codes to perform the analysis presented in Section~\ref{sec:SM}. We extend our thanks to Professor Helmut Eberl (ÖAW) for his valuable comments about the dispersion theoretic calculation of the $h\rightarrow Z\gamma$ amplitude. This work is supported by the research grant \textit{SIGP 400-156.012-014 (GA313-BP-2024)} \textit{Observables de alta precisión en la física del bosón de Higgs}, from the call \textit{Convocatoria Interna de Banco de Proyectos - Año 2024 - Universidad de Pamplona}. \\ \\ \textit{Dedicated to the memory of Susana Partricia Enríquez Ugalde (Oct. 22, 1988 – Feb. 15, 2025), whose support and presence profoundly shaped my journey. Wherever you are now, may you be well and at peace.}


\vspace{5.9mm}
\begin{thebibliography}{999}

\bibitem{B.Abi}
B. Abi et al. Muon g - 2 Collaboration. 
Phys. Rev. Lett. 126, 141801, {\bf 2021}. [arXiv:2104.03281 [hep-ex]]. 
DOI: https://doi.org/10.1103/PhysRevLett.126.141801

\bibitem{CDF}
CDF Collaboration. Science 376, 6589, 170-176, {\bf 2022}. 
DOI: 10.1126/science.abk1781

\bibitem{CMSmW}
CMS Collaboration. CERN-EP-2024-308, {\bf 2024}. [arXiv: 2412.13872 [hep-ex]].
DOI: https://doi.org/10.48550/arXiv.2412.13872

\bibitem{HZgATLASCMS}
G. Aad et al. ATLAS and CMS. 
Phys. Rev. Lett. 132, no.2, 021803, {\bf 2024}. [arXiv:2309.03501 [hep-ex]].
DOI: https://doi.org/10.1103/PhysRevLett.132.021803

\bibitem{ATLASlatest}
ATLAS collaboration. CERN-EP-2025-155, {\bf 2025}. 
[arXiv: 2507.12598 [hep-ex]]. DOI: https://doi.org/10.48550/arXiv.2507.12598

\bibitem{HL-LHC}
ATLAS, CMS Collaborations. ATL-PHYS-PUB-2025-018 CMS-HIG-25-002, {\bf 2025}. 
[arXiv:2504.00672 [hep-ex]]. DOI: https://doi.org/10.48550/arXiv.2504.00672

\bibitem{muontheory1}
S. Kuberski. PoS LATTICE2023, 125, {\bf 2024}. [arXiv:2312.13753 [hep-lat]]. 
DOI: https://doi.org/10.22323/1.453.0125

\bibitem{muontheory2}
A. Boccaletti, S. Borsanyi, M. Davier et al. {\bf 2024}. [arXiv:2407.10913 [hep-lat]].
DOI: https://doi.org/10.48550/arXiv.2407.10913

\bibitem{muontheory3}
J. Bijnens, N. H. Truedsson and A. R. Sánchez. 
JHEP 03, 094, {\bf 2025}. [arXiv:2411.09578 [hep-ph]]. 
DOI: https://doi.org/10.1007/JHEP03(2025)094

\bibitem{muontheory4}
D. G. Melo P., E. A. Reyes R. and A. R. Fazio. 
Particles 7, 2, 327-381, {\bf 2024}. [arXiv:2306.05661 [hep-ph]].
DOI: https://doi.org/10.3390/particles7020020

\bibitem{muontheory5}
T. Aoyama et al. 
Phys. Rep. 887, 1–166, {\bf 2020}. [arXiv:2006.04822 [hep-ph]]. 
DOI: https://doi.org/10.1016/j.physrep.2020.07.006

\bibitem{muontheory6}
M. Davier, A. Hoecker, A.M. Lutz, B. Malaescu, Z. Zhang. 
Eur. Phys. J. C 84, 7, 721, {\bf 2024}. [arXiv:2312.02053 [hep-ph]]. 
DOI: https://doi.org/10.1140/epjc/s10052-024-12964-7

\bibitem{Ellis2024}
J. Ellis, K. A. Olive and V. C. Spanos. 
Eur. Phys. J. C 84, 1121, {\bf 2024}. [arXiv:2407.08679 [hep-ph]]. 
DOI: https://doi.org/10.1140/epjc/s10052-024-13499-7

\bibitem{ATLAS1}
G. Aadet al. ATLAS. 
Eur. Phys. J. C 80, 2, 123, {\bf 2020}. [arXiv:1908.08215 [hep-ex]]. 
DOI: https://doi.org/10.1140/epjc/s10052-019-7594-6

\bibitem{CMS1} 
A. M. Sirunyan et al. CMS. 
JHEP 04, 123, {\bf 2021}. [arXiv:2012.08600 [hep-ex]]. 
DOI: https://doi.org/10.1007/JHEP04(2021)123

\bibitem{PDG}
S. Navas et al. Particle Data Group. Phys. Rev. D 110, 030001, {\bf 2024}. 
DOI: https://doi.org/10.1103/PhysRevD.110.030001

\bibitem{muonWP25}
R. Aliberti et al. [arXiv:2505.21476 [hep-ph]]. 
DOI: https://doi.org/10.48550/arXiv.2505.21476

\bibitem{fnal25}
D. P. Aguillard et al. [arXiv:2506.03069 [hep-ex]]. 
https://doi.org/10.48550/arXiv.2506.03069

\bibitem{MW1}
S. Heinemeyer, W. Hollik, D. Stockinger, A. M. Weber and G. Weiglein. 
JHEP 08, 052, {\bf 2006}. [arXiv:0604147 [hep-ph]]. 
DOI: https://doi.org/10.1088/1126-6708/2006/08/052

\bibitem{MW2}
S. Heinemeyer, W. Hollik, A. M. Weber and G. Weiglein. 
JHEP 04, 039, {\bf 2008}. [arXiv:0710.2972 [hep-ph]]. 
https://doi.org/10.1088/1126-6708/2008/04/039

\bibitem{MW3}
S. Heinemeyer, W. Hollik, G. Weiglein and L. Zeune. 
JHEP 12, 084, {\bf 2013}. [arXiv:1311.1663 [hep-ph]].
DOI: https://doi.org/10.1007/JHEP12(2013)084

\bibitem{DM1}
N. Aghanim et al. Planck. Astron. Astrophys. 641, A6, {\bf 2020}. 
[erratum: Astron. Astrophys. 652, C4, {\bf 2021}]. [arXiv:1807.06209 [astro-ph.CO]]. 
DOI: https://doi.org/10.1051/0004-6361/201833910

\bibitem{DM2}
E. Aprile et al. XENON. 
Phys. Rev. Lett. 121, no.11, 111302, {\bf 2018}. [arXiv:1805.12562 [astro-ph.CO]]. 
DOI: https://doi.org/10.1103/PhysRevLett.121.111302

\bibitem{DM3}
D. S. Akerib et al. LUX. 
Phys. Rev. Lett. 118, no.2, 021303, {\bf 2017}. [arXiv:1608.07648 [astro-ph.CO]]. 
DOI: https://doi.org/10.1103/PhysRevLett.118.021303

\bibitem{DM4}
X. Cui et al. PandaX-II. 
Phys. Rev. Lett. 119, no.18, 181302, {\bf 2017}. [arXiv:1708.06917 [astro-ph.CO]]. 
DOI: https://doi.org/10.1103/PhysRevLett.119.181302

\bibitem{MhExp}
ATLAS, CMS collaboration. 
JHEP 08, {\bf 2016}, 045. [arXiv:1606.02266 [hep-ex]]. 
DOI: https://doi.org/10.1007/JHEP08(2016)045

\bibitem{MW4}
E. Bagnaschi, M. Chakraborti, S. Heinemeyer, I. Saha and G. Weiglein. 
Eur. Phys. J. C 82, 5, 474, {\bf 2022}. [arXiv:2203.15710 [hep-ph]]. 
DOI: https://doi.org/10.1140/epjc/s10052-022-10402-0

\bibitem{FCC1}
P. Azzi et al. {\bf 2012}. [arXiv:1208.1662 [hep-ex]]. 
DOI: https://doi.org/10.48550/arXiv.1208.1662

\bibitem{FCC2}
A. Freitas et al. {\bf 2019}. [arXiv:1906.05379 [hep-ph]]. 
DOI: https://doi.org/10.48550/arXiv.1906.05379

\bibitem{FCC3}
S. Heinemeyer, S. Jadach and J. Reuter. 
Eur. Phys. J. Plus 136, 9, 911, {\bf 2021}. [arXiv:2106.11802 [hep-ph]]. 
DOI: https://doi.org/10.1140/epjp/s13360-021-01875-1

\bibitem{DMhATLAS}
Georges Aad et al. ATLAS Collaboration. 
Phys. Lett. B 847, 138315, {\bf 2023}. [arXiv:2308.07216 [hep-ex]]. 
DOI: https://doi.org/10.1016/j.physletb.2023.138315

\bibitem{DMhCMS}
A. Hayrapetyan et al. CMS Collaboration. Phys. Rev. D 111, 092014, {\bf 2025}. 
[arXiv:2409.13663 [hep-ex]].
DOI: https://doi.org/10.1103/PhysRevD.111.092014


\bibitem{HiggsMass1}
S. P. Martin. 
Phys. Rev. D 106, 1, 013007, {\bf 2022}. [arXiv:2203.05042 [hep-ph]]. 
DOI: https://doi.org/10.1103/PhysRevD.106.013007

\bibitem{HiggsMass2}
E. A. Reyes R. and A. R. Fazio. 
Phys. Rev. D 108, 5, 053007, {\bf 2023}. [arXiv:2301.00076 [hep-ph]]. 
DOI: https://doi.org/10.1103/PhysRevD.108.053007

\bibitem{EWfit}
J. Haller, A. Hoecker, R. Kogler, K. Mönig, T. Peiffer and J. Stelzer.
Eur. Phys. J. C 78, 8, 675, {\bf 2018}. [arXiv:1803.01853 [hep-ph]]. 
DOI: https://doi.org/10.1140/epjc/s10052-018-6131-3

\bibitem{Passarino:2013}
G. Passarino. 
Phys. Lett. B 727, {\bf 2013}, 424-431. [arXiv: 1308.0422 [hep-ph]]. 
DOI: https://doi.org/10.1016/j.physletb.2013.10.052

\bibitem{data}
E. A. Reyes R., C. A. Lopez A., O. R. Torrijo G. and D. G. Melo P.
Ancillary files for ”Rare Higgs Decay into a Photon and a Z Boson in Radiatively-Driven Natural Supersymmetry”. {\bf 2025}. 
\url{https://github.com/fisicateoricaUDP/HiggsDecays.git}


\bibitem{djouadiqcd}
M. Spira, A. Djouadi, and P. M. Zerwas. Phys. Lett. B 276, 350, {\bf 1992}. 
DOI: https://doi.org/10.1016/0370-2693(92)90331-W

\bibitem{numerical}
T. Gehrmann, S. Guns, and D. Kara. JHEP 09, 038, {\bf 2015}. [arXiv:1505.00561 [hep-ph]]. 
DOI: https://doi.org/10.1007/JHEP09(2015)038

\bibitem{zdecay}
R. N. Cahn, M. S. Chanowitz, and N. Fleishon. Phys. Lett. B 82, 113,  {\bf 1979}. 
DOI: https://doi.org/10.1016/0370-2693(79)90438-6

\bibitem{dispersion}
I. Boradjiev, E. Christova and H. Eberl. Phys. Rev. D 97, 073008, {\bf 2018}.
[arXiv:1711.07298 [hep-ph]]. 
DOI: https://doi.org/10.1103/PhysRevD.97.073008

\bibitem{anatomy}
M. Herrero and R. A. Morales. Phys. Rev. D 102, 075040, {\bf 2020}.
[arXiv:2005.03537 [hep-ph]]. 
DOI: https://doi.org/10.1103/PhysRevD.102.075040

\bibitem{2photonsdis}
K. Melnikov and A. Vainshtein. Phys. Rev. D 93, 053015, {\bf 2016}.
[arXiv:1601.00406 [hep-ph]]. 
DOI: https://doi.org/10.1103/PhysRevD.93.053015

\bibitem{NPBgauge}
Tai Tsun Wu and Sau Lan Wu. Nucl. Phys. B 914, 421, {\bf 2017}.
DOI: https://doi.org/10.1016/j.nuclphysb.2016.11.007


\bibitem{FeynCalc1}
V. Shtabovenko, R. Mertig, and F. Orellana. 
Comput. Phys. Commun. 256, 107478 {\bf 2020}.
[arXiv:2001.04407 [hep-ph]]. DOI: https://doi.org/10.1016/j.cpc.2020.107478

\bibitem{FeynCalc2}
R. Mertig, M. Bohm, and A. Denner. 
Comput. Phys. Commun. 64, 345, {\bf 1991}. 
DOI: https://doi.org/10.1016/0010-4655(91)90130-D

\bibitem{FeynArts}
T. Hahn. Comput. Phys. Commun. 140, 418, {\bf 2001}.
[arXiv:0012260 [hep-ph]]. DOI: https://doi.org/10.1016/S0010-4655(01)00290-9

\bibitem{FeynHelpers}
V. Shtabovenko. Comput. Phys. Commun. 218, 48-65, {\bf 2017}. 
[arXiv:1611.06793v2 [physics.comp-ph]]. 
DOI: https://doi.org/10.1016/j.cpc.2017.04.014

\bibitem{PackageX}
H. H. Patel. Comput. Phys. Commun. 197, 276–290, {\bf 2015}.
[arXiv:1503.01469 [hep-ph]]. 
DOI: https://doi.org/10.1016/j.cpc.2015.08.017


\bibitem{hzyrevisited}
A. Djouadi, V. Driesen, W. Hollik. Eur. Phys. J. C 1, 163, {\bf 1998}.
[arXiv:9701342 [hep-ph]]. 
DOI: https://doi.org/10.1007/BF01245806

\bibitem{Yennie}
H. M. Fried and D. R. Yennie. Phys. Rev. 112, 1391, {\bf 1958}. 
DOI: https://doi.org/10.1103/PhysRev.112.1391

\bibitem{generaloneloop}
L. T. Hue, A. B. Arbuzov, T. T. Hong, T. P. Nguyen, D. T.
Si, and H. N. Long. Eur. Phys. J. C 78, 885, {\bf 2018}.
[arXiv:1712.05234 [hep-ph]]. 
DOI: https://doi.org/10.1140/epjc/s10052-018-6349-0

\bibitem{induced}
L. Bergstrom and G. Hulth. Nucl. Phys. B 259, 137
(1985); Nucl. Phys. B276, 744(E), {\bf 1986}. 
DOI: https://doi.org/10.1016/0550-3213(85)90302-5

\bibitem{hunter}
J. F. Gunion, H. E. Haber, G. L. Kane, and S. Dawson. 
Front. Phys. 80, 1, {\bf 2000}. 
DOI: https://doi.org/10.1201/9780429496448

\bibitem{EllisDjouadi}
F. Buccioni, F. Devoto, A. Djouadi, J. Ellis, J. Quevillon, L. Tancredi. 
Phys. Lett. B 851, {\bf 2024}, 138596. [arXiv: 2312.12384 [hep-ph]]. 
DOI: https://doi.org/10.1016/j.physletb.2024.138596

\bibitem{EWNLO1}
Zi-Qiang Chen, Long-Bin Chen, Cong-Feng Qiao and Ruilin Zhu. 
Phys. Rev. D 110, L051301, {\bf 2024}. [arXiv:2404.11441 [hep-ph]]. 
DOI: https://doi.org/10.1103/PhysRevD.110.L051301

\bibitem{EWNLO2}
Wen-Long Sang, Feng Feng, and Yu Jia. 
Phys. Rev. D 110, L051302, {\bf 2024}. [arXiv:2405.03464 [hep-ph]]. 
DOI: https://doi.org/10.1103/PhysRevD.110.L051302


\bibitem{HZPCMS}
A. Tumasyan et al. CMS. 
JHEP 05, 233, {\bf 2023}. [arXiv:2204.12945 [hep-ex]]. 
DOI: https://doi.org/10.1007/JHEP05(2023)233

\bibitem{DMSeeSaw}
N. Das, T. Jha and D. Nanda. 
Phys. Rev. D 109, {\bf 2024}. [arXiv:2402.01317 [hep-ph]]. 
DOI: https://doi.org/10.1103/PhysRevD.109.115020

\bibitem{DHiggsC}
F. Chen, Q. Wen and F. Xu. {\bf 2024}. [arXiv:2412.18540 [hep-ph]]. 
DOI: https://doi.org/10.48550/arXiv.2412.18540

\bibitem{CPViolation}
A. Hernandez-Juarez. R. Gaitan and R. Martinez. 
Phys. Rev. D 111, 015001, {\bf 2025}. [arXiv:2405.03094 [hep-ph]]. 
DOI: https://doi.org/10.1103/PhysRevD.111.015001



\bibitem{hzgSUSY}
J. Cao, L. Wu, P. Wu and J. M. Yang.
JHEP 09, 043, {\bf 2013}. [arXiv:1301.4641 [hep-ph]]. 
DOI: https://doi.org/10.1007/JHEP09(2013)043

\bibitem{MSSM1}
P. Fayet. Nucl. Phys. B 90, 104, {\bf 1975}. ibid. Phys. Lett. B 64, 159, {\bf 1976}. 
ibid. Phys. Lett. B 69, 489, {\bf 1977}. ibid. Phys. Lett. B 84, 416 {\bf 1979}. 
DOI: https://doi.org/10.1016/0550-3213(75)90636-7

\bibitem{MSSM2}
K. Inoue, A. Komatsu and S. Takeshita. 
Prog. Theor. Phys 68, 927, {\bf 1982}. (E) ibid. 70, 330, {\bf 1983}. 
DOI: https://doi.org/10.1143/PTP.68.927

\bibitem{NHMSSM}
S. Israr and M. Rehman. 
Eur. Phys. J. Plus 140, 5, 397, {\bf 2025}. [arXiv:2407.01210 [hep-ph]]. 
DOI: https://doi.org/10.1140/epjp/s13360-025-06401-1

\bibitem{Baer2012}
H. Baer, V. Barger, P. Huang, A. Mustafayev, X. Tata.
Phys. Rev. Lett. 109, 161802, {\bf 2012}. [arXiv:1207.3343 [hep-ph]]. 
DOI: https://doi.org/10.1103/PhysRevLett.109.161802

\bibitem{Baer2013}
H. Baer, V. Barger, P. Huang, D. Mickelson, A. Mustafayev, X. Tata.
Phys. Rev. D 87, 11, 115028, {\bf 2013}. [arXiv:1212.2655 [hep-ph]]. 
DOI: https://doi.org/10.1103/PhysRevD.87.115028

\bibitem{NextMSSM}
U. Ellwanger, C. Hugonie and A. M. Teixeira. 
Phys. Rept. 496, 1, {\bf 2010}. [arXiv:0910.1785 [hep-ph]]. 
DOI: https://doi.org/10.1016/j.physrep.2010.07.001

\bibitem{CMSSM}
G. L. Kane, C. F. Kolda, L. Roszkowski and J. D. Wells. 
Phys. Rev. D 49, 6173, {\bf 1994}. [arXiv:9312272 [hep-ph]]. 
DOI: https://doi.org/10.1103/PhysRevD.49.6173

\bibitem{nMSSM}
M. Maniatis. Int. J. Mod. Phys. A 25, 3505, {\bf 2010}. [arXiv:0906.0777 [hep-ph]]. 
DOI: https://doi.org/10.1142/S0217751X10049827

\bibitem{BLSSM}
A. Hammad, S. Khalil and S. Moretti. 
Phys. Rev. D 92, 9, 095008, {\bf 2015}. [arXiv:1503.05408 [hep-ph]]. 
DOI: https://doi.org/10.1103/PhysRevD.92.095008

\bibitem{uvSSM}
C. X. Liu, H. B. Zhang, J. L. Yang, S. M. Zhao, Y. B. Liu and T. F. Feng. 
JHEP 04, 002, {\bf 2020}. [arXiv:2002.04370 [hep-ph]]. 
DOI: https://doi.org/10.1007/JHEP04(2020)002

\bibitem{BaerJHEP2012}
H. Baer, V. Barger, P. Huang, and X. Tata. 
JHEP 05, 109, {\bf 2012}. 
DOI: https://doi.org/10.1007/JHEP05(2012)109

\bibitem{HiggsSUSY}
P. Slavich et al. 
Eur. Phys. J. C 81, 5, 450, {\bf 2021}. [arXiv:2012.15629 [hep-ph]]. 
DOI: https://doi.org/10.1140/epjc/s10052-021-09198-2

\bibitem{HiggsParticles}
E. A. Reyes R. and A. R. Fazio. 
Particles 5, 1, 53-73, {\bf 2022}. [arXiv:2112.15295 [hep-ph]]. 
DOI: https://doi.org/10.3390/particles5010006

\bibitem{Isasugra}
H. Baer, J. Ferrandis, S. Kraml, and W. Porod. 
Phys. Rev. D 73, 015010, {\bf 2006}. [arXiv:0511123 [hep-ph]]. 
DOI: https://doi.org/10.1103/PhysRevD.73.015010

\bibitem{Isajet}
H. Baer, F. Paige, S. Protopopescu, and X. Tata. 
ISAJET. {\bf 2003}. [arXiv:0312045 [hep-ph]]. 
DOI: https://doi.org/10.48550/arXiv.hep-ph/0312045

\bibitem{FeynHiggs}
H. Bahl, T. Hahn, S. Heinemeyer, W. Hollik, S. Paßehr, H. Rzehak and G. Weiglein. 
Comput. Phys. Commun. 249, 107099, {\bf 2020}. [arXiv:1811.09073 [hep-ph]].
DOI: https://doi.org/10.1016/j.cpc.2019.107099

\bibitem{Baer2025}
H. Baer, V. Barger, J. Bolich and K. Zhang. 
Phys. Rev. D 111, 9, 095019, {\bf 2025}. [arXiv:2412.15356 [hep-ph]]. 
DOI: https://doi.org/10.1103/PhysRevD.111.095019

\bibitem{ILCwhitepaper}
D.~M.~Asner, T.~Barklow, C.~Calancha, K.~Fujii, N.~Graf, H.~E.~Haber, A.~Ishikawa, S.~Kanemura, S.~Kawada and M.~Kurata, \textit{et al.}
{\bf 2018}. [arXiv:1310.0763 [hep-ph]]. 
DOI: https://doi.org/10.48550/arXiv.1310.0763

\bibitem{FCCeeHiggs}
Abada, A., Abbrescia, M., AbdusSalam, S.S. et al. FCC-ee: The Lepton Collider. 
Eur. Phys. J. Spec. Top. 228, 261–623, {\bf 2019}. 
DOI: https://doi.org/10.1140/epjst/e2019-900045-4


\end{thebibliography}
\end{document}